\documentclass[12pt]{article}

\author{Yu.~M.~Zinoviev
       \thanks{E-mail address: Yurii.Zinoviev@ihep.ru} \\
        {\it Institute for High Energy Physics} \\
        {\it Protvino, Moscow Region, 142280, Russia}}
\title{Gravitational cubic interactions \\
for a massive mixed-symmetry gauge field}

\date{}

\def\eptwo{\left\{ \phantom{|}^{\mu\nu}_{ab} \right\}}
\def\epthree{\left\{ \phantom{|}^{\mu\nu\alpha}_{abc} \right\}}
\def\epfour{\left\{ \phantom{|}^{\mu\nu\alpha\beta}_{abcd} \right\}}
\def\epfive{\left\{ \phantom{|}^{\mu\nu\alpha\beta\gamma}_{abcde}%
\right\}}

\begin{document}

\maketitle

\begin{abstract}
In a recent paper \cite{BSZ11} cubic gravitational interactions for a 
massless mixed symmetry field in $AdS$ space have been constructed. In
the current paper we extend these results to the case of massive field.
We work in a Fradkin-Vasiliev approach and use frame-like gauge
invariant description for massive field which works in $(A)dS$ spaces
with arbitrary values of cosmological constant including flat
Minkowski space. In this, massless limit in $AdS$ space coincides with
the results of \cite{BSZ11} while we show that it is impossible to
switch on gravitational interaction for massless field in $dS$ space.
\end{abstract}

\thispagestyle{empty}
\newpage
\setcounter{page}{1}

\section*{Introduction}

In a recent paper \cite{BSZ11} cubic gravitational interactions for
simplest mixed symmetry field (hook) have been investigated using a
number of different approaches, namely:
\begin{itemize}
\item direct construction off all possible cubic vertices using
modified 1 and 1/2 order formalism (similar construction for spin 3
cubic vertices see \cite{Zin10});
\item Fradkin-Vasiliev approach \cite{FV87,FV87a} applied to
Alkalaev-Shaynkman-Vasiliev (ASV) description of massless hook in
$AdS$ space \cite{ASV03,ASV05,ASV06};
\item Fradkin-Vasiliev approach applied to Stueckelberg description of
hook \cite{Zin03a,Zin08c} that differs from the ASV one by the
presence of some Stueckelberg fields;
\item cohomological approach \cite{BH93,Hen98} applied to Stueckelberg
description.
\end{itemize}
While the results of different approaches completely agree, it turns
out that the most simple and straightforward way to construct
interactions is to use Fradkin-Vasiliev approach that initially was
formulated for investigation of gravitational interactions for
massless partialces in $AdS$ space \cite{FV87,FV87a} and then was
successfully applied to more general interactions (see e.g. \
\cite{Vas01,AV02,Alk10}). Let us briefly recall main steps of the
procedure.
\begin{itemize}
\item We begin with frame-like gauge invariant formulation with the
known set of fields and gauge transformations. For each field (both
physical and auxiliary) we construct gauge invariant object that we
will generically call curvature.
\item Then we rewrite free Lagrangian as an expression quadratic in
these gauge invariant curvatures. In general such expression will
contain higher derivatives terms so we have to adjust coefficients so
that all such terms cancel.
\item Now we add quadratic corrections to free curvatures supplemented
with appropriate corrections to gauge transformations so that
variations of deformed curvatures were proportional to the free ones.
\item At last we replace free curvatures in the Lagrangian with the
deformed ones and adjust coefficients so that all variations vanish on
shell. This in turn means that off shell all variations can be
compensated by additional corrections to gauge transformations.
\end{itemize}
As it is clear from this description two main ingredients of such
approach are gauge invariance and frame-like formalism 
\cite{Vas80,LV88,Vas88}. But during a few last years we have seen that
there exists frame-like gauge invariant description for massive fields
both symmetric \cite{Zin08b,PV10} as well as for mixed symmetry ones
\cite{Zin03a,Zin08c,Zin09b,Zin09c}. Moreover such description nicely
works both in flat Minkowski space as well as in $(A)dS$ space with
arbitrary value of cosmological constant, including all possible
massless and partially massless limits. Thus it seems natural to use
Fradkin-Vasiliev approach applied to frame-like gauge invariant
description for investigations of possible interactions for massive
and/or massless particles. In our recent paper \cite{Zin10a}
we have shown how such procedure works in the case of electromagnetic
interactions of massive hook, while the aim of the current paper is to
extend the results of \cite{BSZ11} to the case of massive hook.

In the next section we will give all necessary information on the
free hook including free Lagrangian, gauge transformations and gauge
invariant curvatures. Moreover we will show that using partial gauge
fixing one can obtain a simple description for massive hook directly
related with ASV description for massless one. One of the lessons from
\cite{BSZ11} is that at least for the particular hook case such
partial gauge fixing "commute" with the switching on an interaction so
we may freely use it to simplify calculations without any lost of
generality. Then in Section 2 we consider application of the general
procedure to the gravitational interactions for massive hook including
massless limit in $AdS$ space, while the investigation of massless
case in $dS$ space which turns out to be special moved into Appendix.

{\bf Notations and conventions.} We work in $(A)dS$ space
with $d \ge 4$ dimensions. We will use notation $e_\mu{}^a$ for
background (non-dynamical) frame of $(A)dS$ space and $D_\mu$ for
$(A)dS$ covariant derivatives normalized so that
$$
[ D_\mu, D_\nu ] \xi^a = - \kappa e_{[\mu}{}^a \xi_{\nu]}, \qquad
\kappa = \frac{2\lambda}{(d-1)(d-2)}
$$
We use Greek letters for world indices and Latin letters for local
ones. Surely, using frame $e_\mu{}^a$ and its inverse $e^\mu{}_a$ one
can freely convert world indices into local ones and vice-verse and we
indeed will use such conversion whenever convenient. But separation of
world and local indices plays very important role in a frame-like
formalism. In particular, all terms in the Lagrangians can be written
as a product of forms, i.e. as expressions completely antisymmetric on
world indices and this property greatly simplifies all calculations.
For that purpose we will often use notations $\eptwo = e^\mu{}_a
e^\nu_b - e^\nu{}_a e^\mu{}_b$ and so on.

\section{Kinematics}

Frame-like gauge invariant description \cite{Zin03a,Zin08c} requires
four pairs of physical and auxiliary fields: $(\Omega_\mu{}^{abc},
\Phi_{\mu\nu}{}^a)$, $(\Omega_\mu{}^{ab}, f_\mu{}^a)$, $(C^{abc},
C_{\mu\nu})$ and $(B^{ab}, B_\mu)$. Free Lagrangian describing massive
particle in $(A)dS$ space has the form:
\begin{eqnarray}
{\cal L}_0 &=& - \frac{3}{4} \eptwo \Omega_\mu{}^{acd}
\Omega_\nu{}^{bcd} + \frac{1}{4} \epfour \Omega_\mu{}^{abc}
D_\nu \Phi_{\alpha\beta}{}^d + \nonumber \\
 && + \frac{1}{2} \eptwo \Omega_\mu{}^{ac} \Omega_\nu{}^{bc} -
\frac{1}{2} \epthree \Omega_\mu{}^{ab} D_\nu f_\alpha{}^c - \nonumber
\\
 && - \frac{1}{6} C_{abc}{}^2 + \frac{1}{6} \epthree
C^{abc} D_\mu C_{\nu\alpha} + \frac{1}{4} B_{ab}{}^2 - \frac{1}{2}
\eptwo  B^{ab} D_\mu B_\nu + \nonumber \\
 && + m_1 [ \eptwo \Omega_\mu{}^{abc} f_\nu{}^c + \epthree
\Omega_\mu{}^{ab} \Phi_{\nu\alpha}{}^c ] + \nonumber \\
 && + m_2 [ \epthree \Omega_\mu{}^{abc} C_{\nu\alpha} + \eptwo
C^{abc}  \Phi_{\mu\nu}{}^c] + \nonumber \\
 && + 2 \tilde{m}_2 [ \eptwo \Omega_\mu{}^{ab} B_\nu + e^\mu{}_a 
B^{ab} f_\mu{}^b] + \tilde{m}_1 \eptwo B^{ab} C_{\mu\nu} 
\end{eqnarray}
Here
$$
8 m_1{}^2 - 24 m_2{}^2 = - 3 (d-3) \kappa, \qquad
\tilde{m}_{1,2} = \sqrt{\frac{(d-2)}{(d-3)}}m_{1,2}
$$
The Lagrangian is invariant under the following set of gauge
transformations:
\begin{eqnarray}
\delta_0 \Phi_{\mu\nu}{}^a &=& D_{[\mu} z_{\nu]}{}^a + 
\eta_{\mu\nu}{}^a + \frac{2m_1}{3(d-3)} e_{[\mu}{}^a
\xi_{\nu]} + \frac{4m_2}{(d-3)} e_{[\mu}{}^a \zeta_{\nu]} \nonumber \\
\delta_0 \Omega_\mu{}^{abc} &=& D_\mu \eta^{abc} + \frac{4m_1}{3(d-3)}
e_\mu{}^{[a} \eta^{bc]} \nonumber \\
\delta_0 f_\mu{}^a &=& D_\mu \xi^a + \eta_\mu{}^a + 4 m_1 z_\mu{}^a +
\frac{4\tilde{m}_2}{(d-2)} e_\mu{}^a \Lambda \\ 
\delta_0 \Omega_\mu{}^{ab} &=& D_\mu \eta^{ab} - 2 m_1 \eta_\mu{}^{ab}
\nonumber \\
\delta_0 C_{\mu\nu} &=& D_{[\mu} \zeta_{\nu]} - 2 m_2 z_{[\mu\nu]},
\qquad \delta_1 C^{abc} = 6 m_2 \eta^{abc} \nonumber \\
\delta_0 B_\mu &=& D_\mu \Lambda + 2 \tilde{m}_2 \xi_\mu + 4
\tilde{m}_1 \zeta_\mu, \qquad \delta_0 B^{ab} = - 4 \tilde{m}_2
\eta^{ab} \nonumber
\end{eqnarray}

As the relation on the parameters $m_{1,2}$ clearly shows for non-zero
values of cosmological constant $\kappa$ it is not possible to set
both $m_1$ and $m_2$ equal to zero simultaneously. In $AdS$ space
($\kappa < 0$) one can set $m_2 = 0$. In this, the whole system
decomposes into two disconnected subsystems. One of them with the
Lagrangian and gauge transformations:
\begin{eqnarray}
{\cal L}_0 &=& - \frac{3}{4} \left\{ \phantom{|}^{\mu\nu}_{ab}
\right\} \Omega_\mu{}^{acd} \Omega_\nu{}^{bcd} + \frac{1}{4} \left\{
\phantom{|}^{\mu\nu\alpha\beta}_{abcd} \right\} \Omega_\mu{}^{abc}
D_\nu \Phi_{\alpha\beta}{}^d + \nonumber \\
 && + \frac{1}{2} \left\{ \phantom{|}^{\mu\nu}_{ab} \right\}
\Omega_\mu{}^{ac} \Omega_\nu{}^{bc} - \frac{1}{2} \left\{
\phantom{|}^{\mu\nu\alpha}_{abc} \right\} \Omega_\mu{}^{ab} D_\nu
f_\alpha{}^c \nonumber - \\
 && + m_1 [ \left\{ \phantom{|}^{\mu\nu}_{ab} \right\}
\Omega_\mu{}^{abc} f_\nu{}^c + \left\{
\phantom{|}^{\mu\nu\alpha}_{abc} \right\}
\Omega_\mu{}^{ab} \Phi_{\nu\alpha}{}^c ] 
\end{eqnarray}
\begin{eqnarray}
\delta_0 \Phi_{\mu\nu}{}^a &=& D_{[\mu} z_{\nu]}{}^a + 
\eta_{\mu\nu}{}^a + \frac{2m_1}{3(d-3)} e_{[\mu}{}^a
\xi_{\nu]} \nonumber \\
\delta_0 \Omega_\mu{}^{abc} &=& D_\mu \eta^{abc} + \frac{4m_1}{3(d-3)}
e_\mu{}^{[a} \eta^{bc]} \nonumber \\
\delta_0 f_\mu{}^a &=& D_\mu \xi^a + \eta_\mu{}^a + 4 m_1 z_\mu{}^a \\
\delta_0 \Omega_\mu{}^{ab} &=& D_\mu \eta^{ab} - 2 m_1 \eta_\mu{}^{ab}
\nonumber
\end{eqnarray}
corresponds to massless representation of $AdS$ group (which differs
from that of Poincare group \cite{BMV00,BIS08,BIS08a}), while the
other one just gives gauge invariant description of massive
antisymmetric second rank tensor. In turn, in $dS$ space one can set
$m_1 = 0$. In this case the whole system also decomposes into two
disconnected subsystems. One of them with the Lagrangian and gauge
transformations:
\begin{eqnarray}
{\cal L}_0 &=& - \frac{3}{4} \left\{ \phantom{|}^{\mu\nu}_{ab}
\right\} \Omega_\mu{}^{acd} \Omega_\nu{}^{bcd} + \frac{1}{4} \left\{
\phantom{|}^{\mu\nu\alpha\beta}_{abcd} \right\} \Omega_\mu{}^{abc}
D_\nu \Phi_{\alpha\beta}{}^d + \nonumber \\
 && - \frac{1}{6} C_{abc}{}^2 + \frac{1}{6} \left\{
\phantom{|}^{\mu\nu\alpha}_{abc} \right\} C^{abc} D_\mu
C_{\nu\alpha} + \nonumber \\
 && + m_2 [ \left\{ \phantom{|}^{\mu\nu\alpha}_{abc} \right\}
\Omega_\mu{}^{abc} C_{\nu\alpha} + \left\{
\phantom{|}^{\mu\nu}_{ab} \right\} C^{abc} 
\Phi_{\mu\nu}{}^c]
\end{eqnarray}
\begin{eqnarray}
\delta_0 \Phi_{\mu\nu}{}^a &=& D_{[\mu} z_{\nu]}{}^a + 
\eta_{\mu\nu}{}^a + \frac{4m_2}{(d-3)} e_{[\mu}{}^a \zeta_{\nu]},
\qquad \delta_0 \Omega_\mu{}^{abc} = D_\mu \eta^{abc} \nonumber \\
\delta_0 C_{\mu\nu} &=& D_{[\mu} \zeta_{\nu]} - 2 m_2 z_{[\mu\nu]},
\qquad \delta_1 C^{abc} = 6 m_2 \eta^{abc}
\end{eqnarray}
corresponds to massless representation of $dS$ group, while the other
one describes a so called partially massless spin 2 particle.

In our recent paper \cite{Zin10a} we have investigated electromagnetic
interactions for the same massive mixed symmetry field. We have shown
that it is impossible to take a limit $m_1 \to 0$ without switching
off minimal e/m interactions, while nothing prevents one from taking a
limit $m_2 \to 0$. It turns out that the situation with gravitational
interactions is the same. Namely, in a very recent paper \cite{BSZ11}
cubic gravitational interactions for the case $m_2 = 0$ where
constructed, while in Appendix A of the current paper we consider the
case $m_1 = 0$ and show that it is impossible to switch on
gravitational interaction. Thus in the rest of the paper we will
always assume that $m_1 \ne 0$.

Let us return to the general massive case. Having in our disposal
explicit form of the gauge transformations we can construct gauge
invariant objects (curvatures) for all eight fields (both physical and
auxiliary):
\begin{eqnarray}
{\cal R}_{\mu\nu}{}^{abc} &=& D_{[\mu} \Omega_{\nu]}{}^{abc} +
\frac{4m_1}{3(d-3)} e_{[\mu}{}^{[a} \Omega_{\nu]}{}^{bc]} + 
\frac{4m_2}{3(d-3)} e_{[\mu}{}^{[a} C_{\nu]}{}^{bc]} \nonumber \\
{\cal T}_{\mu\nu\alpha}{}^a &=& D_{[\mu} \Phi_{\nu\alpha]}{}^a -
\Omega_{[\mu,\nu\alpha]}{}^a + \frac{2m_1}{3(d-3)} e_{[\mu}{}^a
f_{\nu,\alpha]}  + \frac{4m_2}{(d-3)} e_{[\mu}{}^a C_{\nu\alpha]}
\nonumber \\
{\cal F}_{\mu\nu}{}^{ab} &=& D_{[\mu} \Omega_{\nu]}{}^{ab} + 2 m_1
\Omega_{[\mu,\nu]}{}^{ab} - \frac{2\tilde{m}_2}{(d-2)}  
e_{[\mu}{}^{[a} B_{\nu]}{}^{b]} \nonumber \\
T_{\mu\nu}{}^a &=& D_{[\mu} f_{\nu]}{}^a - \Omega_{[\mu,\nu]}{}^a - 4
m_1 \Phi_{\mu\nu}{}^a + \frac{4\tilde{m}_2}{(d-2)} e_{[\mu}{}^a
B_{\nu]} \\
{\cal C}_\mu{}^{abc} &=& D_\mu C^{abc} - 6 m_2 \Omega_\mu{}^{abc} - 
\frac{2\tilde{m}_1}{(d-2)} e_\mu{}^{[a} B^{bc]} \nonumber \\
{\cal C}_{\mu\nu\alpha} &=& D_{[\mu} C_{\nu\alpha]} - C_{\mu\nu\alpha}
+ 2 m_2 \Phi_{[\mu\nu,\alpha]} \nonumber \\
{\cal B}_\mu{}^{ab} &=& D_\mu B^{ab} + 4 \tilde{m}_2 \Omega_\mu{}^{ab}
+ \frac{4\tilde{m}_1}{3} C_\mu{}^{ab} \nonumber \\
{\cal B}_{\mu\nu} &=& D_{[\mu} B_{\nu]} - B_{\mu\nu} - 2\tilde{m}_2
f_{[\mu,\nu]} - 4 \tilde{m}_1 C_{\mu\nu} \nonumber
\end{eqnarray}

Now let us partially gauge fix such description by settling $f_\mu{}^a
= 0$ and $B_\mu = 0$. At the same time we solve corresponding
algebraic equations $T_{\mu\nu}{}^a = 0$ and ${\cal B}_{\mu\nu} = 0$:
$$
\Phi_{\mu\nu}{}^a = - \frac{1}{4m_1} \Omega_{[\mu,\nu]}{}^a, \qquad
C_{\mu\nu} = - \frac{1}{4\tilde{m}_1} B_{\mu\nu}
$$
Then (after some field rescaling) we obtain the following simple
Lagrangian for remaining 4 fields:
\begin{eqnarray}
{\cal L}_0 &=& - \frac{3}{4} \eptwo \Omega_\mu{}^{acd}
\Omega_\nu{}^{bcd} - \frac{3}{8} \epthree \Omega_\mu{}^{abd}
D_\nu \Omega_\alpha{}^{cd} - \frac{1}{6} C_{abc}{}^2 - \frac{1}{4}
e^\mu{}_a C^{abc} D_\mu B^{bc} - \nonumber \\
 && - m_2 e^\mu{}_a [ \frac{3}{2} \Omega_\mu{}^{abc}
B^{bc} + C^{abc} \Omega_\mu{}^{bc}] - \frac{m_1{}^2}{2} \eptwo
\Omega_\mu{}^{ac} \Omega_\nu{}^{bc} - \frac{\tilde{m}_1{}^2}{4} 
B_{ab}{}^2
\end{eqnarray}
This Lagrangian is invariant under the following remaining gauge
transformations:
\begin{eqnarray}
\delta \Omega_\mu{}^{abc} &=& D_\mu \eta^{abc} + 
\frac{4m_1{}^2}{3(d-3)} e_\mu{}^{[a} \eta^{bc]}, \qquad 
\delta \Omega_\mu{}^{ab} = D_\mu \eta^{ab} - 2
\eta_\mu{}^{ab} \nonumber \\
\delta C^{abc} &=& 6m_2 \eta^{abc}, \qquad
\delta B^{ab} = - 4m_2 \eta^{ab}
\end{eqnarray}
Note that such description turns out to be closely related with 
Alkalaev-Shaynkman-Vasiliev description for mixed symmetry fields
\cite{ASV03} (see also \cite{ASV05,ASV06}). Indeed gauge fields
$\Omega_\mu{}^{abc}$ and $\Omega_\mu{}^{ab}$ (up to different
normalization) correspond to the ASV description of (partially)
massless hook in $AdS$, while zero forms $C^{abc}$ and $B^{ab}$ play
the roles of Stueckelberg fields making them massive.

After partial gauge fixing we have 4 gauge invariant objects:
\begin{eqnarray}
{\cal R}_{\mu\nu}{}^{abc} &=& D_{[\mu} \Omega_{\nu]}{}^{abc} +
\frac{4m_2}{3(d-3)} e_{[\mu}{}^{[a} C_{\nu]}{}^{bc]} +
\frac{4m_1{}^2}{3(d-3)} e_{[\mu}{}^{[a} \Omega_{\nu]}{}^{bc]}
\nonumber \\
{\cal F}_{\mu\nu}{}^{ab} &=& D_{[\mu} \Omega_{\nu]}{}^{ab} + 2
\Omega_{[\mu,\nu]}{}^{ab} - \frac{2m_2}{(d-3)} e_{[\mu}{}^{[a}
B_{\nu]}{}^{b]} \nonumber \\
{\cal C}_\mu{}^{abc} &=& D_\mu C^{abc} - 6m_2 \Omega_\mu{}^{abc} -
\frac{2m_1{}^2}{(d-3)} e_\mu{}^{[a} B^{bc]} \\
{\cal B}_\mu{}^{ab} &=& D_\mu B^{ab} + \frac{4}{3} C_\mu{}^{ab} + 4m_2
\Omega_\mu{}^{ab} \nonumber
\end{eqnarray}

Our next task is to rewrite the free Lagrangian as an expression
quadratic in these gauge invariant curvatures. The most general such
Lagrangian looks as follows:
\begin{eqnarray}
{\cal L}_0 &=& \epfour [ a_1 {\cal R}_{\mu\nu}{}^{abe} 
{\cal R}_{\alpha\beta}{}^{cde} + a_2 {\cal F}_{\mu\nu}{}^{ab} 
{\cal F}_{\alpha\beta}{}^{cd} ] + \epthree [ a_5 
{\cal R}_{\mu\nu}{}^{abd} {\cal B}_\alpha{}^{cd} + a_6 
{\cal F}_{\mu\nu}{}^{ad}  {\cal C}_\alpha{}^{bcd} ] + \nonumber \\
 && + \eptwo [ a_3 {\cal C}_\mu{}^{acd} {\cal C}_\nu{}^{bcd} + a_4
{\cal B}_\mu{}^{ac} {\cal B}_\nu{}^{bc} ]
\end{eqnarray}
We have usual differential identities for curvatures:
\begin{eqnarray}
D_{[\mu} {\cal R}_{\nu\alpha]}{}^{abc} &=& - \frac{4m_2}{3(d-3)}
e_{[\mu}{}^{[a} {\cal C}_{\nu,\alpha]}{}^{bc]} - 
\frac{4m_1{}^2}{3(d-3)} e_{[\mu}{}^{[a} {\cal F}_{\nu\alpha]}{}^{bc]}
\nonumber \\
D_{[\mu} {\cal F}_{\nu\alpha]}{}^{ab} &=& 2 
{\cal R}_{[\mu\nu,\alpha]}{}^{ab} + \frac{2m_2}{(d-3)} e_{[\mu}{}^{[a}
{\cal B}_{\nu,\alpha]}{}^{b]} \nonumber \\
D_{[\mu} {\cal C}_{\nu]}{}^{abc} &=& - 6m_2 {\cal R}_{\mu\nu}{}^{abc}
+ \frac{2m_1{}^2}{(d-3)} e_{[\mu}{}^{[a} {\cal B}_{\nu]}{}^{bc]} \\
D_{[\mu} {\cal B}_{\nu]}{}^{ab} &=& \frac{4}{3} 
{\cal C}_{[\mu,\nu]}{}^{ab} + 4m_2 {\cal F}_{\mu\nu}{}^{ab} \nonumber
\end{eqnarray}
Note that on the solutions of auxiliary fields $\Omega_\mu{}^{abc}$
and $C^{abc}$ we have
\begin{equation}
{\cal F}_{[\mu\nu,\alpha]}{}^a = 0, \qquad {\cal B}_{[\mu,\nu\alpha]}
= 0 \quad \Longrightarrow \quad {\cal R}_{[\mu\nu,\alpha\beta]}{}^a =
0
\end{equation}
Using these differential identities we can obtain the following 4
identities for curvatures squares:
\begin{eqnarray*}
I_1 &=& \epfive D_\mu [ {\cal R}_{\nu\alpha}{}^{abc} 
{\cal F}_{\beta\gamma}{}^{de}] = \\
 &=& \epfour [ 3 {\cal R}_{\mu\nu}{}^{abe} 
{\cal R}_{\alpha\beta}{}^{cde}
- \frac{2(d-4)m_1{}^2}{(d-3)} {\cal F}_{\mu\nu}{}^{ab}
{\cal F}_{\alpha\beta}{}^{cd} ] - \\
 && - \frac{2(d-4)m_2}{(d-3)} \epthree [
6 {\cal R}_{\mu\nu}{}^{abd} {\cal B}_\alpha{}^{cd} + 4
{\cal F}_{\mu\nu}{}^{ad} {\cal C}_\alpha{}^{bcd} ] = 0 \\
I_2 &=& \epfour D_\mu [ {\cal R}_{\nu\alpha}{}^{abe} 
{\cal C}_\beta{}^{cde} ] = \\
 &=& - 3m_2 \epfour {\cal R}_{\mu\nu}{}^{abe} 
{\cal R}_{\alpha\beta}{}^{cde}  - 
\frac{32(d-4)m_2}{3(d-3)} \eptwo {\cal C}_\mu{}^{acd} 
{\cal C}_\nu{}^{bcd} + \\
 && + \frac{4(d-4)m_1{}^2}{3(d-3)} \epthree
[ 3 {\cal R}_{\mu\nu}{}^{abd} {\cal B}_\alpha{}^{cd} - 2 
{\cal F}_{\mu\nu}{}^{ad} {\cal C}_\alpha{}^{bcd} ] = 0 \\
I_3 &=& \epfour D_\mu [ {\cal F}_{\nu\alpha}{}^{ab} 
{\cal B}_\beta{}^{cd}] = \\
 &=& m_2 \epfour {\cal F}_{\mu\nu}{}^{ab} {\cal
F}_{\alpha\beta}{}^{cd}
+ \epthree [ - 2 {\cal R}_{\mu\nu}{}^{abd} {\cal B}_\alpha{}^{cd} +
\frac{4}{3} {\cal F}_{\mu\nu}{}^{ad} {\cal C}_\alpha{}^{bcd} ] + 8m_2
\eptwo {\cal B}_\mu{}^{ac} {\cal B}_\nu{}^{bc} = 0 \\
I_4 &=& \epthree D_\mu [ {\cal C}_\nu{}^{abd} {\cal B}_\alpha{}^{cd} ]
= \\
 &=& \epthree [ - 3m_2 {\cal R}_{\mu\nu}{}^{abd} {\cal
B}_\alpha{}^{cd} -
2m_2 {\cal F}_{\mu\nu}{}^{ad} {\cal C}_\alpha{}^{bcd} ] + \eptwo [ -
\frac{8}{3} {\cal C}_\mu{}^{acd} {\cal C}_\nu{}^{bcd} + 4m_1{}^2 
{\cal B}_\mu{}^{ac} {\cal B}_\nu{}^{bc} ] = 0
\end{eqnarray*}
These 4 identities are not independent. Indeed by direct calculations
one can show:
$$
m_2 X_1 + X_2 + \frac{2(d-4)m_1{}^2}{(d-3)} X_3 - 
\frac{4(d-4)m_2}{(d-3)} X_4 = 0
$$
Thus we have 3 independent identities so if we require that the
Lagrangian quadratic in curvatures correctly reproduce free Lagrangian
for massive hook given above we would expect that we obtain solution
with 3 arbitrary parameters. This turns out to be the case. Note
however that one has to be careful using this freedom because as our
previous experience shows switching on an interaction tends to
partially resolve this ambiguity. We will use the following simple
choice for free Lagrangian:
\begin{equation}
{\cal L}_0 = \epfour [ a_1 {\cal R}_{\mu\nu}{}^{abe} 
{\cal R}_{\alpha\beta}{}^{cde} + a_2 {\cal F}_{\mu\nu}{}^{ab} 
{\cal F}_{\alpha\beta}{}^{cd} ] + \eptwo [ a_3 {\cal C}_\mu{}^{acd}
{\cal C}_\nu{}^{bcd} + a_4 {\cal B}_\mu{}^{ac} {\cal B}_\nu{}^{bc} ]
\end{equation}
$$
a_1 = - \frac{9}{512m_1{}^2}, \qquad
a_2 = - \frac{3}{256(d-3)}, \qquad
a_3 = - \frac{(d-4)}{16m_1{}^2(d-3)}, \qquad
a_4 = - \frac{3}{32(d-3)}
$$
Later on we will see that such choice is compatible with the
possibility to switch on an interaction.

\section{Cubic gravitational interactions}

For gravitational field we will use notations $h_\mu{}^a$ and
$\omega_\mu{}^{ab}$. Gauge transformations for the free massless field
in $(A)dS$ space have the form:
\begin{equation}
\delta_0 h_\mu{}^a = D_\mu \chi^a + \chi_\mu{}^a, \qquad
\delta_0 \omega_\mu{}^{ab} = D_\mu \chi^{ab} + \kappa
e_\mu{}^{[a} \chi^{b]}
\end{equation}
Correspondingly we have two gauge invariant objects (linearized
curvature and torsion):
\begin{eqnarray}
R_{\mu\nu}{}^{ab} &=& D_{[\mu} \omega_{\nu]}{}^{ab} + \kappa
e_{[\mu}{}^{[a} h_{\nu]}{}^{b]} \nonumber \\
T_{\mu\nu}{}^a &=& D_{[\mu} h_{\nu]}{}^a - \omega_{[\mu,\nu]}{}^a
\end{eqnarray}
For non-zero values of cosmological constant the free Lagrangian can
be written as follows:
\begin{equation}
{\cal L}_0 = - \frac{1}{32\kappa(d-3)} \epfour R_{\mu\nu}{}^{ab}
R_{\alpha\beta}{}^{cd}
\end{equation}

According to general procedure our first task is to find deformations
for all gauge invariant curvatures supplemented with appropriate
corrections to gauge transformations such that variations of these
deformed curvatures were proportional to the free ones.

{\bf Deformations for hooks curvatures.}
Let us consider deformations for hooks curvatures corresponding to
minimal gravitational interactions:
\begin{eqnarray}
\Delta {\cal R}_{\mu\nu}{}^{abc} &=& \omega_{[\mu}{}^{d[a}
\Omega_{\nu]}{}^{bc]d} - \frac{4m_2}{3(d-3)} [ 
h_{[\mu}{}^{[a} C_{\nu]}{}^{bc]} + e_{[\mu}{}^{[a} C^{bc]d} 
h_{\nu]}{}^d ] - \frac{4m_1{}^2}{3(d-3)} h_{[\mu}{}^{[a} 
\Omega_{\nu]}{}^{bc]} \nonumber \\
\Delta {\cal F}_{\mu\nu}{}^{ab} &=& - \omega_{[\mu}{}^{c[a}
\Omega_{\nu]}{}^{b]c} - 2 \Omega_{[\mu}{}^{abc} 
h_{\nu]}{}^c + \frac{2m_2}{(d-3)} [ h_{[\mu}{}^{[a} B_{\nu]}{}^{b]} -
e_{[\mu}{}^{[a} B^{b]c} h_{\nu]}{}^c ] \label {curh} \\
\Delta {\cal C}_\mu{}^{abc} &=& \omega_{[\mu}{}^{d[a}
C^{bc]d} + \frac{2m_1{}^2}{(d-3)} h_\mu{}^{[a} B^{bc]}, \qquad
\Delta {\cal B}_\mu{}^{ab} = - \omega_\mu{}^{c[a} B^{b]c} -
\frac{4}{3} C^{abc} h_\mu{}^c \nonumber
\end{eqnarray}
Similarly, the appropriate corrections to gauge transformations turn
out to be:
\begin{eqnarray}
\delta \Omega_\mu{}^{abc} &=& - \chi^{d[a} \Omega_\mu{}^{bc]d} +
\frac{4m_2}{3(d-3)} [ C_\mu{}^{[ab} \chi^{c]} + e_\mu{}^{[a} C^{bc]d}
\chi^d ] + \frac{4m_1{}^2}{3(d-3)} \Omega_\mu{}^{[ab} \chi^{c]} +
\nonumber \\
 && + \omega_\mu{}^{d[a} \eta^{bc]d} - \frac{4m_1{}^2}{3(d-3)}
h_\mu{}^{[a} \eta^{bc]} \\
\delta \Omega_\mu{}^{ab} &=& \chi^{c[a} \Omega_\mu{}^{b]c} - 2
\Omega_\mu{}^{abc} \chi^c + \frac{2m_2}{(d-3)} [ B_\mu{}^{[a}
\chi^{b]} - e_\mu{}^{[a} B^{b]c} \chi^c ] - \omega_\mu{}^{c[a}
\eta^{b]c} + 2 \eta^{abc} h_\mu{}^c \nonumber \\
\delta C^{abc} &=& - \chi^{d[a} C^{bc]d} - \frac{2m_1{}^2}{(d-3)}
B^{[ab} \chi^{c]}, \qquad \delta B^{ab} = \chi^{c[a} B^{b]c} +
\frac{4}{3} C^{abc} \chi^c \nonumber
\end{eqnarray}
Taking into account these corrections we obtain the following
transformations of deformed curvatures under the hook's 
$\eta^{abc}$ and $\eta^{ab}$ transformations:
\begin{eqnarray}
\delta \hat{\cal R}_{\mu\nu}{}^{abc} &=& R_{\mu\nu}{}^{d[a}
\eta^{bc]d} - \frac{4m_1{}^2}{3(d-3)} T_{\mu\nu}{}^{[a}
\eta^{bc]} \nonumber \\
\delta \hat{\cal F}_{\mu\nu}{}^{ab} &=& 2 \eta^{abc} T_{\mu\nu}{}^c -
R_{\mu\nu}{}^{c[a} \eta^{b]c}
\end{eqnarray}

{\bf Deformations of gravitational curvatures.}
The most general ansatz for such deformations quadratic in fields looks
like (schematically)%
\footnote{Really we have considered the general case without partial
gauge fixing where all eight fields are present. In this, the
resulting expressions for deformed curvatures contain auxiliary fields
$\Omega_3$, $\Omega_2$, $C$ and $B$ only. Thus, at least in this
particular case, partial gauge fixing "commute" with switching on
interactions and we may use it to simplify calculations without any
lost of generality.}:
\begin{eqnarray*}
\hat{R} &\sim& R \oplus \Omega_3 \Omega_3 \oplus \Omega_3 B \oplus B B
\oplus \Omega_2 \Omega_2 \oplus \Omega_2 C \oplus C C \\
\hat{T} &\sim& T \oplus \Omega_3 \Omega_2 \oplus \Omega_3 C
\oplus \Omega_2 B \oplus C B
\end{eqnarray*}
where $\Omega_3$ stands for $\Omega_\mu{}^{abc}$ and $\Omega_2$ ---
for $\Omega_\mu{}^{ab}$. Due to the presence of zero forms $C^{abc}$
and $B^{ab}$ there exists a possibility to make field redefinitions of
the form:
\begin{eqnarray*}
\omega_\mu{}^{ab} &\Rightarrow& \omega_\mu{}^{ab} + \kappa_1 C^{abc}
B_\mu{}^c + \kappa_2 C_\mu{}^{c[a} B^{b]c} + \kappa_3 e_\mu{}^{[a}
C^{b]cd} B^{cd} \\
h_\mu{}^a &\Rightarrow& h_\mu{}^a + \kappa_4 C^{abc} C_\mu{}^{bc} +
\kappa_5 e_\mu{}^a C^{bcd} C^{bcd} + \kappa_6 B^{ab} B_\mu{}^b +
\kappa_7 e_\mu{}^a B^{bc} B^{bc}
\end{eqnarray*}
which we will use to simplify all subsequent expressions%
\footnote{Note that the choice we make here has to be in agreement
with choice for the parameters in the free Lagrangian. As we will see
later on our choices are indeed consistent.}.
In this, the resulting expressions for curvatures can be casted into
the form:
\begin{eqnarray}
\hat{R}_{\mu\nu}{}^{ab} &=& R_{\mu\nu}{}^{ab} + a_0 [ 
\Omega_{[\mu}{}^{acd} \Omega_{\nu]}{}^{bcd} + \frac{m_2}{(d-3)}
e_{[\mu}{}^{[a} \Omega_{\nu]}{}^{b]cd} B^{cd} + \frac{4}{9(d-3)}
C_{[\mu}{}^{ca} C_{\nu]}{}^{bc} - \nonumber \\
 && - \frac{4m_1{}^2}{3(d-3)} \Omega_{[\mu}{}^{ca} 
\Omega_{\nu]}{}^{bc} - \frac{m_1{}^2}{3(d-3)^2} B_{[\mu}{}^a 
B_{\nu]}{}^b + \frac{m_1{}^2}{3(d-3)^2} e_{[\mu}{}^a e_{\nu]}{}^b
B^{cd} B^{cd} ] \\
\hat{T}_{\mu\nu}{}^a &=& T_{\mu\nu}{}^a + a_0 [ \frac{1}{2}
\Omega_{[\mu}{}^{abc} \Omega_{\nu]}{}^{bc} -  \frac{1}{(d-3)} 
C_{[\mu}{}^{ab} B_{\nu]}{}^b - \frac{1}{6(d-3)} e_{[\mu}{}^a 
C_{\nu]}{}^{bc} B^{bc} ] \label{curg} \nonumber
\end{eqnarray}
while appropriate corrections to gauge transformations take the form:
\begin{eqnarray}
\delta \omega_\mu{}^{ab} &=& - a_0 \eta^{cd[a} \Omega_\mu{}^{b]cd} +
\frac{m_2a_0}{(d-3)} e_\mu{}^{[a} \eta^{b]cd} B^{cd} +
\frac{4m_1{}^2a_0}{3(d-3)} \eta^{c[a} \Omega_\mu{}^{b]c}  \nonumber \\
\delta h_\mu{}^a &=& \frac{a_0}{2} [ \Omega_\mu{}^{abc} \eta^{bc} -
\eta^{abc} \Omega_\mu{}^{bc} ]
\end{eqnarray}
For what follows we will need transformations for deformed Riemann
tensor under the hook's $\eta^{abc}$ and $\eta^{ab}$ transformations:
\begin{equation}
\delta \hat{R}_{\mu\nu}{}^{ab} = - a_0 \eta^{cd[a} 
{\cal R}_{\mu\nu}{}^{b]cd} + \frac{4m_1{}^2a_0}{3(d-3)} \eta^{c[a}
{\cal F}_{\mu\nu}{}^{b]c} \label{gvar}
\end{equation}

{\bf Gravitational interaction.} Now according to general procedure we
consider the sum of free Lagrangians for hook and graviton where all
curvatures are replaced with the deformed ones:
\begin{eqnarray}
{\cal L}_0 &=& \epfour [ a_1 \hat{\cal R}_{\mu\nu}{}^{abe} 
\hat{\cal R}_{\alpha\beta}{}^{cde} + a_2 \hat{\cal F}_{\mu\nu}{}^{ab} 
\hat{\cal F}_{\alpha\beta}{}^{cd} ] + \eptwo [ a_3 
\hat{\cal C}_\mu{}^{acd} \hat{\cal C}_\nu{}^{bcd} + a_4 
\hat{\cal B}_\mu{}^{ac} \hat{\cal B}_\nu{}^{bc} ] - \nonumber \\
 && - \frac{1}{32\kappa(d-3)} \epfour \hat{R}_{\mu\nu}{}^{ab}
\hat{R}_{\alpha\beta}{}^{cd} \label{lagint}
\end{eqnarray}
Now we have to consider all variations that do not vanish on shell and
try to adjust coefficients so that all them vanish. Transformations
for hook curvatures we have to take care on look like:
\begin{equation}
\delta \hat{\cal R}_{\mu\nu}{}^{abc} = R_{\mu\nu}{}^{d[a} \eta^{bc]d},
\qquad \delta \hat{\cal F}_{\mu\nu}{}^{ab} = - R_{\mu\nu}{}^{c[a}
\eta^{b]c}
\end{equation}
while for deformed Riemann tensor they are given in (\ref{gvar}).

Variations under the $\eta^{abc}$ transformations give us:
$$
- \epfour [ 4a_1 {\cal R}_{\mu\nu}{}^{abe} 
R_{\alpha\beta}{}^{cf} \eta^{def} + \frac{a_0}{8\kappa(d-3)} 
{\cal R}_{\mu\nu}{}^{aef} R_{\alpha\beta}{}^{bc} \eta^{def} ]
$$
Using on shell identities ${\cal R}_{[\mu\nu,\alpha\beta]}{}^a = 0$
and $R_{[\mu\nu,\alpha]}{}^a = 0$ one can show that the following
identity holds:
$$
\epfour [ 2 {\cal R}_{\mu\nu}{}^{abe} R_{\alpha\beta}{}^{cf} - 
{\cal R}_{\mu\nu}{}^{aef} R_{\alpha\beta}{}^{bc} ] \eta^{def} = 0
$$
Thus we have to put:
\begin{equation}
a_1 = - \frac{a_0}{16\kappa(d-3)} \label{a0}
\end{equation}
Similarly, variations under the $\eta^{ab}$ transformations produce:
$$
\epfour [ 4a_2 {\cal F}_{\mu\nu}{}^{ab} R_{\alpha\beta}{}^{ce}
\eta^{de} - \frac{m_1{}^2a_0}{6\kappa(d-3)^2} 
{\cal F}_{\mu\nu}{}^{ae} R_{\alpha\beta}{}^{bc} \eta^{de} ]
$$
Again using on shell identities ${\cal F}_{[\mu\nu,\alpha]}{}^a = 0$
and $R_{[\mu\nu,\alpha]}{}^a = 0$ one can show that the following
identity holds:
$$
\epfour [ {\cal F}_{\mu\nu}{}^{ab} R_{\alpha\beta}{}^{ce} +
{\cal F}_{\mu\nu}{}^{ae} R_{\alpha\beta}{}^{bc} ] \eta^{de} = 0
$$
Thus we obtain:
\begin{equation}
a_2 = - \frac{m_1{}^2a_0}{24\kappa(d-3)^2} = 
\frac{2m_1{}^2a_1}{3(d-3)}
\end{equation}
Note that the resulting relation for $a_1$ and $a_2$ is in agreement
with our choice for free Lagrangian.

Thus the Lagrangian (\ref{lagint}) with the deformed curvatures
defined in (\ref{curh}) and (\ref{curg}) gives us a correct set of
cubic gravitational vertices including standard minimal interactions
together with non-minimal higher derivatives ones. There are two
particular limits that one can consider here. First of all we may take
$m_2 = 0$. In this limit Stueckelberg fields $C^{abc}$ and $B^{ab}$
completely decouple and the result (up to different field
normalization) completely agree with the results obtained previously
in \cite{BSZ11}.

Other interesting and important limit is a flat limit $\kappa \to 0$
(i.e. $m_2{}^2 \to m_1{}^2/3$). The peculiarity here is related with
the fact that for massless graviton it is possible to rewrite
Lagrangian as an expression quadratic in curvatures for non-zero
values of cosmological constant only. But from the relation 
(\ref{a0}) we obtain:
\begin{equation}
a_0 = \frac{9(d-3)\kappa}{32m_1{}^2}
\end{equation}
so that at least in the linear approximation the contribution from
gravity part of the Lagrangian is non-singular in a flat limit.

\section*{Conclusion}

Thus we have seen that Fradkin-Vasiliev approach together with
frame-like gauge invariant formalism for massive fields allows one
effectively investigate possible interactions for massive and/or
massless fields. The massive hook (as well as massive spin 2) is one
of the simplest examples possible but it is clear that such approach
can be applied to higher spin fields (both symmetric and mixed
symmetry ones) as well. One of the questions that deserves further
study is the problem of flat limit for gravitational interactions. The
reason is that the Lagrangian for massless graviton can be written as
square of curvature for the non-zero cosmological constant only though
as we have seen in the linear approximation flat limit is
non-singular. Also it would be interesting to understand a striking
difference between massless representations in $AdS$ and $dS$ spaces
as far as switching on interactions is concerned.

\vskip 1cm \noindent
{\bf Acknowledgment}  \\
Author is grateful to N. Boulanger, E. Skvortsov and M. Vasiliev for
many useful discussions. The work was supported in parts by RFBR grant
No.11-02-00814.

\appendix

\section{Partially massless case in a de Sitter space}

Here we will try to switch on gravitational interactions for
(partially) massless mixed symmetry field in $dS$ space corresponding
;to $m_1 \to 0$ limit.

{\bf Kinematics.} For convenience we reproduce here gauge
transformations for this case:
\begin{eqnarray}
\delta_0 \Phi_{\mu\nu}{}^a &=& D_{[\mu} z_{\nu]}{}^a + 
\eta_{\mu\nu}{}^a + \frac{4m_2}{(d-3)} e_{[\mu}{}^a \zeta_{\nu]},
\qquad
\delta_0 \Omega_\mu{}^{abc} = D_\mu \eta^{abc}  \nonumber \\
\delta_0 C_{\mu\nu} &=& D_{[\mu} \zeta_{\nu]} - 2 m_2 z_{[\mu\nu]},
\qquad \delta_0 C^{abc} = 6 m_2 \eta^{abc} 
\end{eqnarray}
Here $8 m_2{}^2 = (d-3) \kappa$. Correspondingly we have four gauge
invariant objects:
\begin{eqnarray}
{\cal R}_{\mu\nu}{}^{abc} &=& D_{[\mu} \Omega_{\nu]}{}^{abc} +
\frac{4m_2}{3(d-3)} e_{[\mu}{}^{[a} C_{\nu]}{}^{bc]} \nonumber \\
{\cal T}_{\mu\nu\alpha}{}^a &=& D_{[\mu} \Phi_{\nu\alpha]}{}^a -
\Omega_{[\mu,\nu\alpha]}{}^a + \frac{4m_2}{(d-3)} e_{[\mu}{}^a
C_{\nu\alpha]} \nonumber \\
{\cal C}_\mu{}^{abc} &=& D_\mu C^{abc} - 6 m_2 \Omega_\mu{}^{abc} \\
{\cal C}_{\mu\nu\alpha} &=& D_{[\mu} C_{\nu\alpha]} - C_{\mu\nu\alpha}
+ 2 m_2 \Phi_{[\mu\nu,\alpha]} \nonumber
\end{eqnarray}
It is not hard to express the free Lagrangian in terms of these gauge
invariant curvatures:
\begin{equation}
{\cal L}_0 = \epfour [ a_1 {\cal R}_{\mu\nu}{}^{abe} 
{\cal R}_{\alpha\beta}{}^{cde} + a_2 \eptwo {\cal C}_\mu{}^{acd} 
{\cal C}_\nu{}^{bcd} + a_3 {\cal T}_{\mu\nu\alpha}{}^a 
{\cal C}_\beta{}^{bcd} ]
\end{equation}
$$
\frac{512m^2(d-4)a_1}{3(d-3)} - 48m^2 a_2 = - 1, \qquad
a_3 = - \frac{1}{72m}
$$
Again there is an ambiguity in the choice of parameters related with
differential identities. Indeed we have:
\begin{eqnarray}
D_{[\mu} {\cal R}_{\nu\alpha]}{}^{abc} &=& - \frac{4m_2}{3(d-3)}
e_{[\mu}{}^{[a} {\cal C}_{\nu,\alpha]}{}^{bc]} \nonumber \\
D_{[\mu} {\cal C}_{\nu]}{}^{abc} &=& - 6m_2 {\cal R}_{\mu\nu}{}^{abc}
\end{eqnarray}
As a result we obtain:
$$
\epfour D_\mu [ {\cal R}_{\nu\alpha}{}^{abe} {\cal C}_\beta{}^{cde} ]
= - 3m_2 [ \epfour {\cal R}_{\mu\nu}{}^{abe} 
{\cal R}_{\alpha\beta}{}^{cde} + \frac{32(d-4)}{9(d-3)} \eptwo
{\cal C}_\mu{}^{acd} {\cal C}_\nu{}^{bcd} ] = 0
$$

{\bf Deformations for gravitational curvatures.}
This time, due to the presence of zero form, there is an ambiguity
related with possible field redefinitions, namely:
$$
h_\mu{}^a \Longrightarrow h_\mu{}^a + \alpha_1 C_\mu{}^{bc} C^{abc} +
\alpha_2 e_\mu{}^a C^{bcd} C^{bcd}
$$
Using these redefinitions deformed curvatures can be casted into the
form:
\begin{eqnarray}
\hat{R}_{\mu\nu}{}^{ab} &=& D_{[\mu} \omega_{\nu]}{}^{ab} +
\kappa e_{[\mu}{}^{[a} h_{\nu]}{}^{b]} + 6m_2 a_0 [ 
\Omega_{[\mu}{}^{acd} \Omega_{\nu]}{}^{bcd} - \frac{4}{3(d-3)}
C_{[\mu}{}^{ac}  C_{\nu]}{}^{bc} ] \nonumber \\
\hat{T}_{\mu\nu}{}^a &=& D_{[\mu} h_{\nu]}{}^a - 
\omega_{[\mu,\nu]}{}^a - a_0 \Omega_{[\mu}{}^{abc} C_{\nu]}{}^{bc} 
\end{eqnarray}
In this, appropriate corrections to gauge transformations look as
follows:
\begin{equation}
\delta \omega_\mu{}^{ab} = - 6m_2a_0 \eta^{cd[a} \Omega_\mu{}^{b]cd},
\qquad \delta h_\mu{}^a = a_0 \eta^{abc} C_\mu{}^{bc} 
\end{equation}
Taking into account these corrections we obtain the following
transformations for deformed curvatures:
\begin{equation}
\delta \hat{R}_{\mu\nu}{}^{ab} = - 6m_2a_0 \eta^{cd[a} 
{\cal R}_{\mu\nu}{}^{b]cd}, \qquad
\delta \hat{T}_{\mu\nu}{}^a = a_0 \eta^{abc} 
{\cal C}_{[\mu,\nu]}{}^{bc}
\end{equation}

{\bf Deformations for hook's curvatures.} In these case the desired
results can be easily obtained by the usual substitutions
corresponding to minimal gravitational interactions:
\begin{eqnarray}
\hat{\cal R}_{\mu\nu}{}^{abc} &=& D_{[\mu} \Omega_{\nu]}{}^{abc} +
\frac{4m_2}{3(d-3)} e_{[\mu}{}^{[a} C_{\nu]}{}^{bc]} + \nonumber \\
 && + \Omega_{[\mu}{}^{d[ab} \omega_{\nu]}{}^{c]d} - 
\frac{4m_2}{3(d-3)} h_{[\mu}{}^{[a} C_{\nu]}{}^{bc]} -
\frac{4m_2}{3(d-3)} e_{[\mu}{}^{[a} C^{bc]d} h_{\nu]}{}^d \nonumber
\\
\hat{\cal T}_{\mu\nu\alpha}{}^a &=& D_{[\mu} \Phi_{\nu\alpha]}{}^a -
\Omega_{[\mu,\nu\alpha]}{}^a + \frac{4m_2}{(d-3)} e_{[\mu}{}^a
C_{\nu\alpha]} - \nonumber \\
 && - \omega_{[\mu}{}^{ab} \Phi_{\nu\alpha]}{}^b -
\Omega_{[\mu,\nu}{}^{ab} h_{\alpha]}{}^b -
\frac{4m_2}{(d-3)} h_{[\mu}{}^a C_{\nu\alpha]} \\
\hat{\cal C}_\mu{}^{abc} &=& D_\mu C^{abc} - 6 m_2 \Omega_\mu{}^{abc}
+ \omega_\mu{}^{d[a} C^{bc]d} \nonumber \\
\hat{\cal C}_{\mu\nu\alpha} &=& D_{[\mu} C_{\nu\alpha]} -
C_{\mu\nu\alpha} + 2 m_2 \Phi_{[\mu\nu,\alpha]}  + h_{[\mu}{}^a
C_{\nu\alpha]}{}^a - 2m_2 \Phi_{[\mu\nu}{}^a h_{\alpha]}{}^a \nonumber
\end{eqnarray}
Corrections to gauge transformations turn out to be:
\begin{eqnarray}
\delta \Omega_\mu{}^{abc} &=& - \chi^{d[a} \Omega_\mu{}^{bc]d} +
\frac{4m_2}{3(d-3)} [ \chi^{[a} C_\mu{}^{bc]} - e_\mu{}^{[a}
C^{bc]d} \chi^d ] - \eta^{d[ab} \omega_\mu{}^{c]d} \nonumber \\
\delta \Phi_{\mu\nu}{}^a &=& \chi^{ab} \Phi_{\mu\nu}{}^b +
\Omega_{[\mu,\nu]}{}^{ab} \chi^b + \frac{4m_2}{(d-3)} \chi^a
C_{\mu\nu} + \eta_{[\mu}{}^{ab} h_{\nu]}{}^b - 
\omega_{[\mu}{}^{ab} z_{\nu]}{}^b - \frac{4m_2}{(d-3)}
h_{[\mu}{}^a \zeta_{\nu]} \nonumber \\
\delta C^{abc} &=& - \chi^{d[a} C^{bc]d}, \qquad 
\delta C_{\mu\nu} =  - \chi^a C_{\mu\nu}{}^a + 2m_2
\Phi_{\mu\nu}{}^a \chi^a + 2m_2 z_{[\mu}{}^a h_{\nu]}{}^a 
\end{eqnarray}
Under the hook's gauge transformations $\eta^{abc}$, $z_\mu{}^a$ and
$\zeta_\mu$ these deformed curvatures transform as follows:
\begin{eqnarray}
\delta \hat{\cal R}_{\mu\nu}{}^{abc} &=& - \eta^{d[ab} 
R_{\mu\nu}{}^{c]d} , \qquad \delta \hat{\cal C}_{\mu\nu\alpha} = -
2m_2 T_{[\mu\nu}{}^a z_{\alpha]}{}^a  \nonumber \\
\delta \hat{\cal T}_{\mu\nu\alpha}{}^a &=& - \eta_{[\mu}{}^{ab}
T_{\nu\alpha]}{}^b - R_{[\mu\nu}{}^{ab} z_{\alpha]}{}^b - 
\frac{4m_2}{3(d-3)} T_{[\mu\nu}{}^a \zeta_{\alpha]} 
\end{eqnarray}

{\bf Gravitational interactions.} Following general procedure we
consider sum of the free Lagrangians for hook and graviton but with
all curvatures replaced with the deformed ones:
\begin{eqnarray}
{\cal L}_0 &=& \epfour [ a_1 \hat{\cal R}_{\mu\nu}{}^{abe} 
\hat{\cal R}_{\alpha\beta}{}^{cde} + a_2 \eptwo 
\hat{\cal C}_\mu{}^{acd} \hat{\cal C}_\nu{}^{bcd} + a_3 
\hat{\cal T}_{\mu\nu\alpha}{}^a \hat{\cal C}_\beta{}^{bcd} ] -
\nonumber \\
 && - \frac{1}{256m_2{}^2} \epfour \hat{R}_{\mu\nu}{}^{ab}
\hat{R}_{\alpha\beta}{}^{cd}
\end{eqnarray}
Now we have to take care on variations that do not vanish on shell:
\begin{equation}
\delta \hat{R}_{\mu\nu}{}^{ab} = - 6m_2a_0 \eta^{cd[a} 
{\cal R}_{\mu\nu}{}^{b]cd}, \quad
\delta \hat{\cal R}_{\mu\nu}{}^{abc} = - \eta^{d[ab} 
R_{\mu\nu}{}^{c]d}, \quad
\delta \hat{\cal T}_{\mu\nu\alpha}{}^a =  
- R_{[\mu\nu}{}^{ab} z_{\alpha]}{}^b  
\end{equation}
and try to adjust coefficients so that all of them vanish. It is easy
to see that this time it is impossible. Crucial point --- $z_\mu{}^a$
transformations that give
$$
\delta {\cal L} \sim \epfour {\cal C}_\mu{}^{abc} R_{\nu\alpha}{}^{de}
z_\beta{}^e
$$
and this can not be compensated even on mass shell!

\end{document}